\documentstyle[aaspp4,12pt,epsf]{article}

\def\micron{\mu m}

\def\kpband{$K'$ band}

\def\mag{\,\hbox{mag}}
\def\arcsec{\hbox{arcsec}}

\def\yr{\,\hbox{yr}}
\def\sec{\,\hbox{s}}
\def\km{\,\hbox{km}}
\def\kpc{\,\hbox{kpc}}
\def\Mpc{\,\hbox{Mpc}}
\def\kms{\,\hbox{km}/\hbox{s}}
\def\CO{\hbox{\rm CO}}
\def\dCO{\delta \hbox{CO}}
\def\DCO{\Delta \hbox{CO}}
\def\COd{{\CO}_\delta}
\def\1CO{{\CO}_1}
\def\2CO{{\CO}_2}
\def\DCO{\Delta \hbox{CO}}

\def\obsCO{{\CO}_{\hbox{obs}}}
\def\trueCO{{\CO}_{\hbox{true}}}
\def\dfc{\delta f_{2.21}}
\def\dfb{\delta f_{2.34}}
\def\fc{f_{2.21}}
\def\fb{f_{2.34}}
\def\muk{\mu_{_{K'}}}
\def\Teff{{\rm T}_{{\rm eff}}}
\def\K{\,{\rm K}}
\def\ADU{\,\hbox{ADU}}
\def\LRV96{Lan\c{c}on, Rocca-Volmerange, \& Thuan (1996)}

\def\fbbar{\overline{\fb}}
\def\Msun{\,M_\odot}
\def\cB{c_B}
\def\cS{c_S}
\def\cY{c_Y}
\def\bB{b_B}
\def\bS{b_S}
\def\bY{b_Y}

\begin{document}
\title{Young red supergiants and the near infrared light
appearance of disk galaxies}

\author{James E. Rhoads\altaffilmark{1}}
\affil{Princeton University Observatory, Peyton Hall, Princeton, NJ
08544 \\
 I: jrhoads@noao.edu}
\authoraddr{Kitt Peak National Observatory, P.O. Box 26732, Tucson, AZ
85726-6732}
\altaffiltext{1}{Present address: Kitt Peak National Observatory, P.O.
Box 26732, Tucson, AZ 85726-6732}

\begin{abstract}
Disk galaxies often show prominent nonaxisymmetric features at
near-infrared wavelengths.  Such features may indicate variations in
the surface density of stellar mass, contributions from young red
supergiants in star forming regions, or substantial dust
obscuration.  To distinguish among these possibilities, we
have searched for spatial variations in the $2.3 \micron$
photometric CO index within the disks of three nearby galaxies (NGC
278, NGC 2649, \& NGC 5713).  This index measures the strength of the
absorption bands of molecular CO in stellar atmospheres, and is strong
in cool, low surface-gravity stars, reaching the largest values for red
supergiants.

We observe significant spatial CO index variations in two galaxies
(NGC 278 \& NGC 5713), indicating that the dominant stellar population
in the near-infrared is not everywhere the same.  Central CO index
peaks are present in two galaxies; these could be due to either
metallicity gradients or recent star formation activity.  In addition,
significant azimuthal CO index variations are seen in NGC 278.
Because strong azimuthal metallicity gradients are physically
implausible in disk galaxies, these features are most naturally
explained by the presence of a young stellar population.

The fraction of $2 \micron$ light due to young stellar populations in
star forming regions can be calculated from our data.  Overall, young
stellar populations can contribute $\sim 3\%$ of a (normal) galaxy's
near infrared flux, which is consistent with other global properties
(e.g., the near-IR Tully-Fisher relation).  Locally, this
fraction may rise to $\sim 33\%$. Thus, young stars do not dominate
the total near infrared flux, but can be locally dominant in star forming
regions, and can bias estimates of spiral arm amplitude or other
nonaxisymmetric structures in galaxies' mass distributions.

\end{abstract}

\section{Introduction}

In the search for reliable tracers of stellar mass,  near infrared
light has become a popular choice.  Astrophysical arguments in its
favor are the relative insensitivity of near infrared light to hot,
young stars, and the relatively small obscuration by dust at these
wavelengths.   The recent development of near infrared array detectors
has  opened the way for surface photometry at $1$--$3.5 \micron$ wavelengths,
allowing practical studies of galaxies' structure in this wavelength
range.

Studies of disk galaxies have shown that spiral arms and other
nonaxisymmetric features frequently have large amplitudes--- of order
2:1 arm to interarm contrast or more (Rix \& Rieke 1993, Rix \&
Zaritsky 1995, Terndrup et al 1994).  Spiral arms are
prominent at blue wavelengths, where hot, luminous OB stars emit much of
their radiation.  Blue light can thus be said
to trace sites of star formation.  Because near infrared (NIR) light in H
($1.6 \mu m$) and K ($2.2 \mu m$) bands is in the Rayleigh-Jeans part
of the spectrum for even very cool stellar photospheres, 
the contributions of hot stars there are minimized
and the longer-lived, cooler stars are well represented.
NIR light is also less attenuated by interstellar dust than is optical
light, which will make arms appear smoother at $2 \micron$ than at
optical wavelengths.

If a galaxy formed most of its stars more than a rotation period
($\sim 10^8$ years) ago, then the older stellar populations have had
time to orbit the galaxy, and will no longer reflect the spatial
distribution of star formation regions.  Assuming these old stars
dominate the near
infrared (NIR) light, this light will reflect the underlying stellar mass
distribution.  If it is strongly concentrated to spiral
arms,  large arm amplitudes must then be explained by stellar dynamics
alone.  This can be restated by noting that if NIR light is dominated
by old stars, then NIR light approximately obeys a continuity
equation.  Nonaxisymmetric features a factor $\alpha$ brighter than
the background disk then imply velocity perturbations of order
$(\alpha-1)/\alpha \times (v_c - \Omega_p R)$, where $v_c$ is the
circular speed, $\Omega_p$ the angular pattern speed, and $R$ the
galactocentric distance.  For a 2:1 arm to interarm ratio ($\alpha=2$)
and regions of the galactic disk away from corotation resonance (where
we expect $|v_c - \Omega_p R| / |v_c| \sim 0.5$), we obtain velocity
perturbations $ \sim v_c / 4$.

However, the contribution of young stars to NIR light is not zero.  In
fact, it can greatly exceed the contribution from the upper main
sequence alone, because some fraction of high mass stars pass through
a luminous red supergiant phase at the ends of their lives. 
These stars are much younger than typical disk galaxy rotation periods
(age $\sim 10^7$ years [e.g., Schaller et al 1992]),
and so will still trace the stellar
birthplaces with reasonable fidelity.  If they contribute a
substantial fraction of the near infrared light from a galaxy, the
advantages of NIR as a measure of stellar mass distribution may be but
little.  In this work, we attempt to directly determine how much
young supergiant stars contribute to features in NIR images of
disk galaxies, and hence how well the stellar mass distribution of
disk galaxies can be inferred from NIR surface photometry.

Determining the amount of supergiant light in the near-infrared has
several applications.  First, orbit synthesis models for the mass
distribution and dynamics of disk galaxies often attempt to reproduce a
surface mass density proportional to NIR surface brightness 
(Athanassoula 1996, Quillen 1996, Grosb\o{}l \& Patsis 1996).
Replacing this direct proportionality with something more accurate
would be beneficial for such models.

Second, population synthesis models that seek to predict the spectra
of composite stellar populations may benefit if substantially improved
constraints on the numbers of supergiants become available.

Third, theoretical models of the late evolution of massive stars are
highly uncertain, primarily because of difficulties modeling mass
loss, convective overshooting, and semiconvection (e.g. Langer \&
Maeder 1995).  The
factors controlling whether a particular star will or will not become
a red supergiant are thus ill-understood.  However, different models
will lead to different predictions for the numbers of red supergiants,
and by studying the contribution of red supergiants to near-IR light
we may provide an observational check on stellar evolution theory.

In the following sections, we will discuss our chosen tool, the CO
index (section~II); describe the observations (section~III) and
analysis (section~IV); present our results (section~V); and discuss
the implications of our findings (section~VI).

\section{2.3 $\micron$ CO absorption as a diagnostic tool}

The $2.3 \micron$ absorption due to CO molecules in the atmospheres of
cool stars provides a useful tool for the study of stellar populations
at near infrared wavelengths.  The strength of this absorption is
traditionally measured using the CO index, which is essentially a
photometric color defined between intermediate bandpass filters at
wavelengths of $2.21 \micron$ and $2.34 \micron$.  Specifically, $\CO =
2.5 \log_{10}\left({F_{2.21} / F_{2.34} } \right) + q $, so
that larger CO indices indicate stronger absorption features.  The
constant $q$ is adjusted so that stars with no CO absorption have
index $\approx 0$.

The CO index increases strongly as surface gravity decreases, which
makes it an effective diagnostic of stellar luminosity class (Baldwin,
Frogel, \& Persson 1973).  It also increases with increasing stellar
metallicity for giant stars of (approximately) fixed surface
temperature (measured by $V-K$ color) (Frogel, Cohen, \& Persson
1983), and increases with decreasing surface temperature (measured by
$J-K'$ color) for Galactic giant stars (McWilliam \& Lambert 1984).
Note that several different filter sets have been used to define CO
indices in the literature (e.g. Baldwin et al 1973, Frogel et al 1978,
McWilliam \& Lambert 1984), and additional definitions have been used
to characterize spectroscopically measured CO absorption (e.g.
Lan\c{c}on, Rocca-Volmerange, \& Thuan 1996).  This can complicate
comparison of CO indices from different sources.  We have set the zero
point of our CO indices using the standard star list of Elias et al
(1982), who use the filter system defined by Frogel et al (1978).  The
CO index of Vega is set to $0$ in this system.  The central
wavelengths and bandpasses of our continuum filter ($2.21 \micron$,
$0.10 \micron$) and CO band filter ($2.34 \micron$, $0.09 \micron$)
are similar to  those used by Frogel et al (1978) ($2.20 \micron$,
$0.11 \micron$; $2.36 \micron$, $0.08 \micron$).  Thus, our indices
should reproduce the system of Frogel et al (1978) fairly well.

The behavior of the CO index for stellar atmosphere models of differing
metallicity, surface
temperature, and surface gravity has been explored in detail by Bell
\& Briley (1991).  The dependences on metallicity and surface
temperature follow na\"{\i}ve intuition well: In order for CO molecules to
be abundant, there must be adequate supplies of carbon and oxygen, and
the temperature must not be so high that the molecules dissociate.
Thus the CO index decreases with decreasing metallicity or increasing
surface temperature.  The dependence on surface gravity is less
obvious.  As pressure increases, the opacity in the CO absorption band
rises, but the continuum opacity rises faster.  The net effect is that
the equivalent width of the absorption bands rises for stars of lower
surface gravity (Bell \& Tripicco 1991; Bell \& Briley 1991).
Frogel (private communication) points out that microturbulence
increases with decreasing surface gravity, and so offers another
mechanism for the luminosity dependence:  As the microturbulent
velocity rises, saturated CO lines are broadened and their equivalent
widths rise.

Because the index measures the continuum only on the blue side of the
absorption feature, it is affected by changes in the continuum slope
even in the absence of CO absorption.  For pure black body spectra,
$CO \approx -0.07 (J - K) \approx -0.15 (H - K)$.  Thus in cool stars
zero index does not exactly correspond to no absorption.

If the CO index shows spatial variations in a galaxy, it implies a
spatial variation of the underlying stellar population.  In general,
this will also imply a variation in the near infrared light to mass
ratio.  Because differential rotation causes phase mixing of the stars
in disk galaxies on timescales of a few rotation periods (i.e., a few
$\times 10^8$ years), while chemical evolution proceeds over the star
forming lifetime of the galaxy (i.e., $\sim 10^{10}$ years for large
disk galaxies), no substantial azimuthal variation in metallicity is
expected.  Azimuthal variations in the CO index are therefore most
naturally explained by star formation activity $ \la 10^8 $ years ago.

\section{Observations}

\subsection{Sample selection}

The galaxies in our sample were face-on disk galaxies chosen to span a
range of spiral arm morphologies and star formation activity.  In
addition, they had to be small enough to fit within the $2 \arcmin$
field of the instrument (described below), and to lie in a suitable
part of the sky for observation from Apache Point Observatory
(latitude $32^\circ 46' 49.3''$ N).

The sample consists of three galaxies.  The first of these, NGC 278,
was chosen because its optical image (Wray 1988; Sandage \& Bedke
1994) shows prominent bright patches in its inner regions.  These are
presumably areas of active star formation, making NGC 278 a good place
to look for the signature of young stars in near infrared light.  It
also has a very high central surface brightness ($\mu_K \approx 14.1
\mag / \arcsec^2$ in $1.3''$ seeing).  It lies at the
lowest redshift in the sample (heliocentric velocity $641 \kms$ [de
Vaucouleurs et al 1991]), meaning that its distance is the least
certain (with perhaps 50\% errors possible due to peculiar velocity).

The second galaxy, NGC 2649, is a grand design spiral with smooth arms
of moderate amplitude.  No nonstellar objects (besides the galaxy
itself) are apparent in our images of NGC 2649, from which we conclude
that it has no massive satellites ($M \ga 5\times 10^8 \Msun \approx
0.003 M(\hbox{{\rm NGC 2649}})$) of normal mass-to-light ratio within
17 kpc.  The Palomar Observatory Sky Survey (POSS) plates allow this
limit to be extended to at least 130 kpc at a slightly less stringent
limit (no satellite with $M \gg 0.02 M(\hbox{{\rm NGC 2649}})$).

Finally, the third galaxy, NGC 5713, has an unusual morphology that
suggests it has recently been disturbed, perhaps by an interaction
with another galaxy.  NGC 5713 has a large FIR to optical luminosity
ratio (twice that of NGC 278 and six times that of NGC 2649), suggesting
a large mean star formation rate.

Properties of these galaxies derived from the literature and from our
photometry are summarized in Table~I.

\tablenum{1}
\begin{table}

\begin{tabular}{|l|ccc|}
\cline{1-4}
Galaxy &	 NGC 2649 & 	NGC 278 & 	NGC 5713 \\
 \cline{1-4}
RA (2000.0) & 	8:44:06.0 & 	 	0:52:01.6 & 	14:40:11.52 \\
Dec (2000.0) &	34:43:01.6 &		47:33:18 &	-0:17:25.8 \\
$V_\odot$ ($\km/\sec$) \tablenotemark{a}
	& $4244 \pm 6$ & $641 \pm 9$ & $1883 \pm 4$ \\
$V_{lg} ($\km/\sec$)  $\tablenotemark{b}
	&$4249 \pm 6$ & 	$931 \pm 9$ &	$1777 \pm 4$ \\
$d$ ($\Mpc$) \tablenotemark{c} 
		& 57	 & 12.4	 & 24 \\
$m_{FIR} - B_T^0 $\tablenotemark{d}
&	0.36 &	-1.02 & 	-1.71 \\
$m_{21}^0 - B_T^0 $\tablenotemark{e}
&	2.82 &	 2.87 & 	1.47  \\
$K'$ \tablenotemark{f} & 10.11 & 7.73 & 8.43 \\  
$J$  \tablenotemark{f} &       &      & 9.50 \\  
$B_T^0 - K'$ & 2.71 & 2.86 & 3.08 \\
$r_{exp}(2.22 \micron)$\tablenotemark{g} & $15.6'' \pm 1.5$
	& $ 11.7'' \pm 0.6 $ & $14'' \pm 1$ \\
 \cline{1-4}
\end{tabular}
\tablenotetext{a}{Heliocentric velocity from de Vaucouleurs et al 1991 [RC3].}
\tablenotetext{b}{Velocity relative to the local group, determined
  using a correction formula supplied by M. Strauss.}
\tablenotetext{c}{Calculated as $ d = V_{lg} / H_0$,
  assuming $H_0=75 \kms / \Mpc$.}
\tablenotetext{d}{Far infrared to extinction- and
  inclination-corrected blue light ratio, in magnitude units, from RC3.}
\tablenotetext{e}{Corrected 21 cm line flux to blue light ratio,  in magnitude
  units, from RC3.}
\tablenotetext{f}{Total near-infrared magnitudes, derived from Apache Point
  surface photometry.  Estimated $1 \sigma$ errors are $0.05 \mag$.  No
  corrections for extinction or inclination have been applied.}
\tablenotetext{g}{Exponential scale length based on $2.22 \micron$ maps.}
\end{table}

\subsection{Observations}

All observations were taken at the Astrophysical Research Consortium (ARC)
3.5 meter telescope at Apache
Point Observatory\footnote{APO is privately
owned and operated by the Astrophysical Research Consortium (ARC),
consisting of the University of Chicago, Institute for Advanced Study,
Johns Hopkins University, New Mexico State University, Princeton University,
University of Washington, and Washington State University.}
, New Mexico, USA, using the facility near-infrared
camera, GRIM II (built by M. Hereld, B. Rauscher, and collaborators).

The data  were taken and reduced following standard practice for near
infrared photometry.  Additionally, the data were processed
independently through a model fitting technique that allows better
control of certain systematic errors; the results of the two methods
were consistent.  Further details of the observations are
given in the remainder of this section.  The data analysis procedures
are described in the following section.

We used three filters: A broad \kpband\ filter (Wainscoat \& Cowie 1992)
with central wavelength $2.114 \micron$ and $0.343 \micron$ full width
at half maximum (FWHM);
a narrow continuum filter with transmission-weighted mean wavelength
$2.210 \micron$ and FWHM $0.100 \micron$ ($4.5\%$); and a narrow CO band
filter with mean wavelength $2.34 \micron$ and FWHM $0.091 \micron$
(3.9\%).  (Quoted filter characteristics are based on filter traces
made by the manufacturer at the operating temperature for the filters.)

The data was taken over four nights, 1995 December 11 (NGC 278, NGC
2649), 1995 December 14 (NGC 2649), 1995 December 31 (NGC 278, NGC
2649), and 1996 April 29 (NGC 5713). 
Conditions were photometric on December 11 and on April 29, while some
cirrus was present on the remaining two nights.  Absolute calibration
of the photometry used only data from the photometric nights. 
Seeing was typically $1$-$1.5''$ on December 11, 14, and 31, and
was $\sim 4''$ on April 29.

On 1996 April 29, we obtained useful J, H, and \kpband\ calibration
data from three stars in the Elias et al (1982)
list (HD 201941,  HD 129655, and HD 136754).  Four stars
(BD+3$^\circ$2954, HD
161903, HD 201941, and HD 129655) were  observed in the CO index
filters.   The airmass range covered was from $1.2$
to $3.9$.  The resulting transformations from instrumental to standard
magnitudes were
$m_{K'} = K - (22.354 \pm 0.014) + (0.079 \pm 0.006) X$ 
and 
$m_{2.34} - m_{2.21} = CO + (0.283 \pm 0.014) + (0.032 \pm 0.006) X$.
Here K, J, and CO are tabulated standard magnitudes, $m_{K'}$,
$m_{2.34}$, and $m_{2.21}$ are instrumental magnitudes defined as
$-2.5 \log_{10}(\ADU/\sec)$, and $X$ is the airmass of the
observation.

On 1995 December 11, useful photometric calibration data was obtained
in the CO index filters from G77-31 and HD 40335 (from the list
by Elias et al [1982]); and in \kpband\ from Feige 11, SA94-242, and
GD71 (from the UKIRT standards list by Casali \& Hawarden [1992]).
The less extensive range of airmass ($1.1$ to $1.6$) and color data
allowed only zero points to be computed.  These were $m_{K'} = K -
(22.191 \pm 0.007)$ and $m_{2.34} - m_{2.21} = CO + (0.373 \pm
0.010)$.  The airmass range of the actual data on 1995 December 11 is
small ($1.05$ to $1.35$), so that uncorrected extinction terms are
unlikely to exceed $0.02 \mag$ in \kpband\ and $0.01 \mag$ in CO
index.

Observations of each source consisted of a series of short (9--40
second) exposures of the object, interspersed with equal length
exposures of nearby sky frames.  These exposures were chosen to
accumulate $\sim 5 \times 10^4$ electrons per pixel from the night
sky, so that read noise ($\sim 100$ electrons) is not a dominant
source of error while the chip remains well below saturation
thresholds ($\sim 13 \times 10^4$ electrons per pixel).  The telescope
pointing was shifted randomly by a few arcseconds between exposures,
so that a given point on the sky was sampled by many different pixels
of the detector.  To extract estimates of the target field's surface
brightness distribution from individual object frames, we first
subtracted the sky background using a combination of two to four sky
frames taken immediately before and after the object frame.  This step
also removed most bias and dark current contribution to the signal.
Because the bias structure of the chip showed small fluctuations from
exposure to exposure and from line to line, the next step was to apply
an additive correction to each line, ensuring that regions of the chip
far from the galaxy have median pixel value zero.  This step also has
the effect of subtracting any residual sky signal, which may be
important when the variable sky brightness is not adequately described
by a linear trend in time.  Note that the galaxies all had $2.21
\micron$ scale lengths $\le 16''$, which is sufficiently small
compared to the $121''$ chip that useful sky level measurements are
possible at the edges of object frames.

The sky-subtracted data frames were then flat fielded, i.e., divided
by a flat field image in order to correct for pixel-to-pixel
variations in the detector's quantum efficiency and gain.  We used
``supersky'' flat field images.  These were constructed for each band
and each night of observations by subtracting a mean dark current
frame from a large number of sky frames, taking a pixel-by-pixel
median of those frames, and normalizing the result to have median
pixel value unity.  

Finally, the individual data frames were registered and combined.
Registration used astrometry of a bright, sharply peaked source (generally
the nucleus of the galaxy, occasionally a nearby star) to determine
appropriate fractional pixel shifts, which were then applied using
bilinear interpolation.  The shifted frames were then combined
using a pixel-by-pixel median.  The median ensures that bad
pixels, cosmic ray hits, stars in the sky frames, and other artifacts
in individual object frames have little effect on the final map, since
such imperfections typically affect $\la 1\%$
of the pixels in a given data frame.

The instrument response is mildly nonlinear.  We have reduced the data
first without and subsequently with nonlinearity corrections.  The
only significant change to the results is in the zero points of the
photometry: CO indices are reduced by $0.05$ in the linearized data,
and total magnitudes of the galaxies change by $0.05$ -- $0.11$.
Relative CO index measurements between different regions of the same
galaxy were essentially unaffected.  This is because the galaxies have
a lower surface brightness than the night sky almost everywhere, so
that the effect of the nonlinearity on sky-subtracted galaxy images is
essentially to change the instrument gain by a few percent from image
to image.  We emphasize that our analysis is based primarily on
azimuthal variations in the CO index, whose significance is unaffected
by zero point shifts of the index.

An alternative data reduction method was used independently to produce
images less sensitive to sky subtraction errors and bias structure
variations in individual exposures.  The method treats each pixel of
the output image as one parameter of a linear model.  Additional
parameters representing bias level offsets in the data and offsets in
the sky level of the image are included in a natural way.  The model
parameters yield a predicted data set through a linear map that can be
determined using the flat field response of the chip and pointing
information derived from astrometry of the individual exposures.  The
linear system is then solved for the model parameters using a
regularized $\chi^2$ minimization (cf. Press et al 1992, sections
18.4--18.5).   Similar techniques have been applied in the
mid infrared by Van Buren \& Kong (1997).  Our algorithm will be
described in greater detail elsewhere.

Reduced $K'$ band images of the galaxies are shown in figures 1a--1c.

\section{Analysis}

The data analysis was done by defining masks that group pixels
according to their surface brightness and location within the galaxy,
and measuring the CO index of each pixel group.
This method, which we call ``masked aperture photometry,'' can combine
the relatively high  signal to noise ratios of traditional
large-aperture photometry with the physical insight of a spatially
resolved map.

Because we want to know what fraction of NIR light is contributed by
young supergiants, we define masks according to the ratio of the
measured surface brightness to a smooth axisymmetric model for the
galaxy. In detail, if
the local surface brightness is given by $f$ and the model by
$m$, we define bright regions by $f/m - 1 > \tau_1$,
and background regions by $f/m - 1 < \tau_2$, with
thresholds $\tau_2 < \tau_1$.  The choice of an appropriate model and
threshold is an important step in the procedure.  We considered using
the axisymmetrized image of the galaxy as our model.  However, this
method will fail to flag the excess surface brightness from star
formation in rings (like that in M31) or nuclei.  We therefore chose
to model the background light with a pure exponential disk.  The
exponential parameters were derived from a fit to the one-dimensional
K' band profiles of the galaxies.  The masks are shown in figures
1d--1f.

For any set ${\cal S}$ of pixels, we can define the total CO index of
${\cal S}$ in the obvious way:
$\CO({\cal S}) = 2.5 \log_{10}\left({
\int_{\cal S}  \fc dA /
 \int_{\cal S} \fb dA } \right) + q $, where $\fc$ and $\fb$ are the
local surface brightness in the $2.21 \micron$ and $2.34 \micron$
filters, $\int_{\cal S} dA$ represents the integral over the solid
angle contained in ${\cal S}$, and $q$ sets the zero-point of the CO
index scale.

Once the masks have been made, we can take their intersections with
annuli about the galaxy center and form a curve of the CO index as a
function of radius for bright and background regions.  This curve can
be made either cumulative or differential.  In this work we favor
the differential version of the CO index- radius curves, since these
have better spatial detail.  The cumulative curves have higher total
signal to noise ratio but offer less insight into where the data reliably
measure differences in CO index.

\subsection{Error estimation}

The error in the measured CO index is composed of both random and
systematic parts.  These are of comparable importance in the data
analysis.

Random errors are due to simple photon counting statistics.  Because
the surface brightness of our sample galaxies is generally well
below that of the NIR night sky (reaching a peak of $1/3$ of sky on
the nucleus of NGC 278), we estimated photon counting noise
from the measured variance of sky pixels in our final images.

The most problematic systematic error is imperfect sky subtraction.
The sky background is very bright (ranging from 12.8 to 13.4
mag/arcsec$^2$ in $K'$ in our data) and can change by $\sim 1\%$ on
timescales of a minute or so.  Because it contains multiple components
(night sky emission lines, thermal emission by the atmosphere, and
thermal emission by the telescope and instrument), 
the flat field response may also vary
slightly.  To overcome this, it is necessary to take sky frames before
and after  object exposures and subtract a running median of the sky
observations from each object frame.  This eliminates most of the sky
flux, and any residuals can be eliminated using a uniform value scaled
by the flat field.  Errors after this last step will be no worse than $ 23$
mag/arcsec$^2$.
In practice, an additional complication was introduced by the fluctuations in
bias level on the chip.  Because bias level offsets are not
proportional to the flat field, they cannot be handled exactly as sky
level errors are.  Distinguishing the amount of each error based on
the outer regions of the chip is not always straightforward.  In the
worst case, the error introduced by handling a bias level offset as
sky or vice versa is about 2 magnitudes worse than the above estimate,
or up to $21$ mag/arcsec$^2$.  These errors will vary in amplitude and
sign from exposure to exposure, so should average out roughly as
$1/\sqrt{N_{\exp}}$.  They can also have some spatial structure,
reflecting structure in the flat field.

We have used the variations in standard star fluxes measured at
different places on the detector on 1996 April 29 to study possible
residual flat fielding errors in our data.  Standard star measurements
that night were taken in an ``X'' pattern, i.e., one central
measurement near the middle of the chip, and four more offset NE, NW,
SE, and SW by $25''$ in right ascension and $25''$ in declination.  We
sky subtracted and flat fielded these data in the normal way, replaced
bad pixels with the median of their neighbors, and measured standard
star fluxes.  We then compared the flux at each offset position with
that at the central position.  The mean magnitude offsets in the four
quadrants were $-0.0073 \pm 0.0075$, $0.0066 \pm 0.0070$,
$0.0027 \pm 0.0065$, and $0.0081 \pm 0.0091$.  We repeated these
calculations for the CO continuum filter alone and for the CO band
filter alone, and found no evidence for systematic differences between
the two filters.  Thus, this test presents no compelling evidence for
flat field errors, and rules out flat field errors much in excess of
$1\%$.  Note also that such errors would affect only the
sky-subtracted flux in the image; that is, a residual $1\%$ flat field
error here means $1\%$ of the object surface brightness and not $1\%$
of the sky level.

Consider now the effect of a spatially constant sky level error on the
CO index measurement.  For the interesting regime (where systematic
errors are smaller than the mean surface brightness), the error $\dCO$
can be derived as follows:
$$ \CO + \dCO = 2.5 \log_{10}\left( {\fc + \dfc \over \fb +
\dfb } \right) + q =  2.5 \log_{10}\left({\fc ( 1 + \dfc/ \fc) \over
\fb (1 + \dfb / \fb) } \right) + q$$
$$ =  2.5 \log_{10}\left( {\fc / \fb } \right) + q
+  2.5 \log_{10}\left(1 + \dfc/ \fc \right)
- 2.5  \log_{10}\left(1 + \dfb/ \fb \right)$$
whence 
$$ \dCO = 2.5 \log_{10}\left(1 + \dfc/ \fc \right)
- 2.5  \log_{10}\left(1 + \dfb/ \fb \right)$$
$$ \approx 1.086 (\dfc / \fc - \dfb  / \fb ) ~~. $$
Although the
magnitude of this error is in general $\sim |\dfc / \fc| \sim | \dfb  /
\fb |$, it can be considerably smaller if $\dfb /\dfc \approx \fb /\fc$.
Such a cancellation may be expected for some sky subtraction
algorithms.  For example, if the sky level is determined in an annulus
around the center of a galaxy, the sky level errors will be
proportional to the galaxy flux at some outer isophote.  When such
cancellation is present,
we can define the quantity $\COd  = 2.5
\log_{10}(\dfc/\dfb) + q$ and derive the approximate relations
$$ \dCO \approx 1.086 \left( {\dfc \over \fc} - {\dfb  \over \fb} \right)
= \left( \dfb \over \fb \right) \times 1.086 \left( {\dfc \over \dfb} { \fb
\over \fc} - 1 \right)$$
$$
 \approx \left( \dfb \over \fb \right) \times 2.5 \log_{10}
\left( { \dfc \over \dfb } {\fb \over \fc} \right)
= { \dfb \over \fb }(\COd - \CO) \approx { \dfc \over \fc }(\COd - \CO)
$$

In our data set, we estimated the sky subtraction errors by examining
exponential fits to the galaxies' radial profiles.  The radial
profiles from our data reduction procedure fall below the best fit
exponentials at radii ranging from $\sim 2.25$ (for NGC 2649) to $\sim
5.5$ scale lengths (for NGC 278).  By adding a suitably chosen
constant, the light profile can be made to follow the exponential fit
to the lowest measurable surface brightness in the data.  This
constant is taken as our estimate of the sky subtraction error in the
individual filters.
Using this estimator, we find $-0.5 \la \COd \la 0.3$ for our
galaxies, and we therefore use the larger of $0.5  (\dfc / \fc)$ and
$0.5  (\dfb / \fb)$ for $\dCO$.
This assumes a modest cancellation of the sky subtraction errors in
the two bands. 
This analysis suggests that sky subtraction errors in the CO index
are $\sim \pm 0.01 $ at $\muk  = 18$ mag/arcsec$^2$, with some
differences from galaxy to galaxy. 
Note that the stellar disks of galaxies are believed to cut off at $\sim 4$
[optical] scale lengths (e.g., Robin, Crez\'{e}, \& Mohan 1992), so that the
observed end of the exponential profile may be real in most of our
sample.  If so, our error estimates are conservative.


Because sky subtraction errors are largely a function of local surface
brightness, the accuracy of a {\sl differential} measurement of CO
index between two regions at similar surface brightness can be much
better than the above calculations suggest.  Consider the quantity
$(\1CO - \2CO) = 2.5 \log_{10}\left[ ({\fc}_{,1} / {\fb}_{,1}) \times
({\fb}_{,2} / {\fc}_{,2}) \right]$ when the surface brightnesses in
regions 1 and 2 are comparable.  In the ideal limit ${\fc}_{,1} =
{\fc}_{,2} = \fc$, $(\1CO - \2CO) = 2.5 \log_{10}\left( {\fb}_{,2} /
{\fb}_{,1} \right) \approx 1.086 \left[ ({\fb}_{,2} - {\fb}_{,1})/
\fbbar \right] $, which is clearly unaffected by errors in $\fc$.
(Here $\fbbar \equiv ({\fb}_{,1} + {\fb}_{,2}) / 2 \approx {\fb}_{,1}
\approx {\fb}_{,2}$).  Errors $\dfb$ affect it as $\delta (\1CO -
\2CO) \approx 1.086 \left( \dfb / {\fb}_{,2} - \dfb / {\fb}_{,1}
\right) \approx 1.086 \, \dfb \left[ ({\fb}_{,1} - {\fb}_{,2}) / \fbbar^2
\right] \approx -(\dfb / \fbbar) \times (\1CO - \2CO)$.
Thus a differential CO index measurement can be accurate to $\pm 0.01$
even in the presence of fractionally large sky subtraction errors.  In
practice, this means relative measurements are good at the $\delta
(\1CO - \2CO) < 0.01$ magnitude level down to $\muk \approx 19.5$ in
our data, provided the regions under comparison are at similar $\muk$
and are subject to identical sky subtraction errors.

A second source of systematic error is seeing variations.  If the
point spread function (PSF) is appreciably broader in one band than another,
artificial color gradients will be induced near strong peaks
(especially point sources) in the image.  The magnitude of this effect
is easily estimated under the approximation of a Gaussian PSF.
Consider a change in the breadth $\sigma$ of the Gaussian PSF by a
factor $(1 + \alpha)$ between two bands.
Then the surface brightness ratio at the peak of a point source will
change by a factor $(1 + \alpha)^2$, which introduces a color shift of
$5 \alpha  / \ln(10)$ magnitudes in the limit of small $\alpha$.  For an
extended source, the effect is diminished.  For a source with a
Gaussian profile of intrinsic width $\eta$, a change from seeing sigma
$\beta$ to $(1 + \alpha) \beta$ introduces a color shift 
$5 \alpha  / \ln(10) / \left[ 1 + (\eta/\beta)^2 \right]$.

To avoid such seeing-induced color shifts, we have matched the spatial
resolution of the 2.21 and 2.34 $\micron$ images for each galaxy.
This was done by convolving the higher resolution image with a
Gaussian kernel of appropriate width.  For 
NGC 278 and NGC 2649, the kernel width was determined by comparing the
widths of stars present in the mosaics.  In  NGC 5713, our
field does not contain stars.  We therefore measured the FWHM of the
galaxy nucleus in each sky-subtracted exposure.  For NGC 5713, the
nucleus FWHM varied smoothly throughout the observations, with no
obvious discontinuities when changing between $2.34$ and $2.21
\micron$ filters (which was done four times in all).  We therefore
smoothed the $2.21 \micron$ image to match the observed width of the
$2.34 \micron $ image. 

Even after this smoothing, NGC 5713 and NGC 278 have significantly
increased CO indices in their nuclei.  Residual errors in PSF
matching could be of order $\alpha \la 0.02$, and the nuclei are not
point sources ($\eta / \beta \approx 1$), so that errors in the central
indices are of order $0.02$.   We therefore conclude that
these two galaxies show a real increase in CO absorption in their nuclei.
Outside the galactic nuclei, where features are less sharply peaked
and our masks cover larger solid angles, the CO indices did not change
measurably when seeing corrections were applied.

A final systematic effect that must be considered is redshift.  The CO
absorption beyond $2.29 \micron$ is composed of many individual bands,
with bandheads separated by about $0.03 \micron$ and a width $\sim
0.0075 \micron$ at half of the maximum absorption.  The $2.34 \micron$
filter contains only a few such features,
and the measured CO index may change
appreciably as these features redshift through the edges of the
bandpass.  At the lowest redshifts, where the entire filter lies
redward of the bandhead, these changes will be small and not
necessarily monotonic in redshift.  At somewhat higher redshifts
($1000 \kms \la ca \la 11000 \kms$), the CO absorption covers only
part of the bandpass, and the correction becomes substantial.  At
redshifts $cz \gg 12000 \kms$ the CO index no longer measures CO
absorption at all.  (These velocity ranges are quoted for the APO
filter set.)  To compensate for this effect, we have calculated
synthetic CO indices for model stellar spectra (from Bell \& Briley
1991) kindly provided by R.  Bell and redshifted to the recessional
velocities of our sample galaxies.  The correction is roughly
proportional to the strength of the absorption, and was calculated for
a range of rest frame CO indices.  The effect of continuum slope on
the CO index must be considered when calculating redshift corrections,
because at $0.003 \la z \la 0.04$, the amount of absorption in the CO
filter changes much more with redshift than the continuum slope does.
Taking the $J-K$
color as indicative of continuum slope, we suggest a redshift
correction equation of the form $ \obsCO = \alpha(z) \left[ \trueCO +
\beta (J - K) \right] - \beta (J - K)$.  Here $\beta \approx 0.07$
(based on measured colors for black body spectra of different
temperatures), and $\alpha(z)$ is determined by integration over model
spectra.

The approximate behavior of $\alpha$ can be modeled by pretending that
the optical depth of CO absorption is zero for $\lambda < 2.29
\micron$, constant for $\lambda > 2.29 \micron$, and similarly that
the filter transmission is uniform in $\lambda_1 < \lambda <
\lambda_2$ and zero elsewhere.  Then $\alpha(z) \approx (\lambda_2 -
(1+z) 2.29 \micron) / (\lambda_2 - \lambda_1)$ for $\lambda_1 < (1+z)
2.29 \micron < \lambda_2$.  Combining this approximation with the
redshift correction formula above gives 
$\trueCO - \obsCO \approx \left[ (1+z) 2.29 \micron - \lambda_1
\right] / (\lambda_2 - \lambda_1) \times 
\left[ \trueCO + \beta ( J-K ) \right]$.  We can
compare this to Frogel et al's (1978) correction, $\trueCO - \obsCO
\approx 4.8z$, by inserting values appropriate to the old stellar
systems they studied and their filter system ($J-K \approx 0.9$,
$\trueCO \approx 0.15$, $\lambda_1 = 2.32 \micron$, $\lambda_2 = 2.40
\micron$).  This yields $\trueCO - \obsCO \approx 6 (z - 0.013)$ for
$0.013 < z < 0.048$, which is fairly good agreement given the
crudeness of our approximation for $\alpha(z)$.

We have corrected our measured CO indices for redshift effects using
the general formula above.  The values of $\alpha$ obtained for our
sample by convolving APO filters with model spectra are $\alpha(641 \kms) =
0.957$, $\alpha(1883 \kms) = 0.868$, and $\alpha(4244 \kms) = 0.645$.
We have measured $J-K = 1.07$ for NGC 5713.  For the other two
galaxies, we use $J-K \approx 0.9$, which is typical for Sb and Sbc
galaxies like these (Peletier \& Balcells 1996).  We use $\beta =
0.07$.

\section{Results and Interpretation}

The CO index (corrected for redshift effects) 
is plotted as a function of angular radius for all sample
galaxies in figure~2.
The degree of spatial variation in observed CO index differs
substantially among
galaxies in our sample.  At one extreme, NGC~278 shows strong and
statistically significant differences ($\DCO \approx 0.07$)
between the CO indices of bright patches and of the background disk in
the annulus $16'' < r < 22''$.  In addition, it shows
a  CO index peak ($\CO \approx 0.14$) in the nucleus. 
The feature at $r \approx 20''$ corresponds to a prominent ring of
star forming regions in the galaxy.  Beyond $25''$, there is no
significant difference between the CO indices of different regions.
There is a radial gradient in measured CO index, but this should be
interpreted with extreme caution, as radial trends are quite sensitive
to sky subtraction problems.

\begin{figure}
\figurenum{2}
\plotone{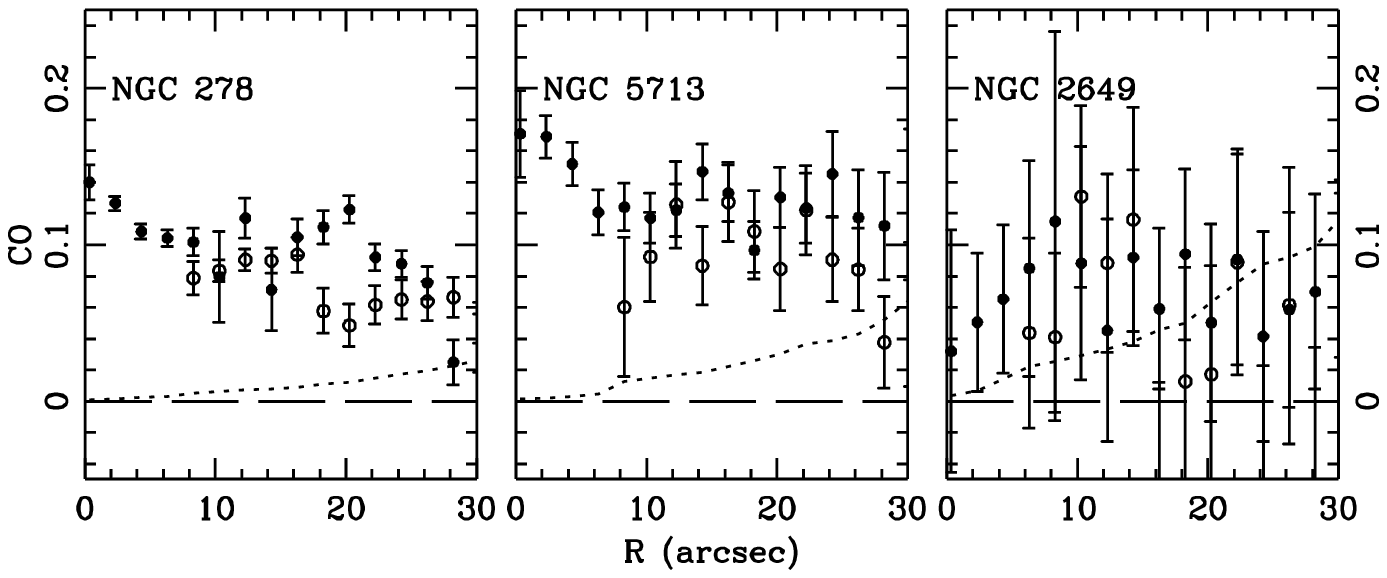}
\caption{
CO index profiles for the sample galaxies, corrected for
redshift effects.  Solid points represent the
CO index of bright regions; open points represent the CO index of
background regions.  Error bars on the individual points represent
random errors due to photon statistics.  The dotted curve represents
the estimated sky subtraction error in the CO index zero point as a
function of radius.  At radii where both bright and background region
CO indices are defined, the sky subtraction error plotted is for
background regions; that for bright regions is typically smaller by a
factor of $1.5$--$2$.  The dashed curve marks the $x$-axis for
reference.  Note that the difference between the CO indices of bright
and background regions is less affected by sky subtraction errors than
is the CO index zero point.  }
\end{figure}

NGC 5713 also shows a central peak in CO absorption at $\CO
\approx 0.17$.   At
larger radii there are weak variations in CO index.  The index of
bright patches lies generally above that for faint regions from $8''$
to $30''$, but with significance rarely exceeding $1 \sigma$.  The
background disk has an index $\CO \approx 0.10$ in the inner regions.
The ``bright region'' mask in this galaxy basically follows the
bulge,  bar, and single bright spiral arm as galactocentric radius
increases.  
The gradient in CO index near the center could be due to either
nuclear star formation activity or a metallicity gradient. 

NGC 2649 shows no significant variations in CO index between regions
or between nucleus and disk.  While this galaxy has the lowest surface
brightness in the sample and therefore modest signal to noise ratio,
arm-interarm differences  $\DCO \ga 0.10$ can be ruled out.  

\subsection{Overall CO indices}
The overall CO indices measured for our sample are generally below those of
elliptical and S0 galaxies published by Frogel et al (1978).  We 
have examined the expected
evolution of CO index with time for a composite stellar population, using
a modified version of Bruzual and Charlot's GISSEL spectral synthesis package
to produce CO index predictions for solar metallicity and various star
formation histories.  These predictions will underestimate the CO
index for periods when supergiants dominate the near-IR light, because
the spectral libraries used were incomplete for M supergiants (Bruzual
\& Charlot 1993) and M giant spectra were used instead.  However, for
older stellar populations (age $t \ga 10^{8.5}$ years) the results
should be fairly accurate.  These results (shown in figure~3) show
only a weak dependence of CO index on population age for a galaxy with
continuous star formation ongoing for $t \ge 10^{8.5} \yr$, with an
overall slight increase in CO index from $10^{8.5} \yr$ to $10^{10.3}
\yr$.  The CO index for an instantaneous burst fluctuates considerably
more but the overall trend is again a weak increase from $10^{8.5}
\yr$ to $10^{10.3} \yr$.  The model index for a typical elliptical
galaxy population (a single burst $\sim 10^{10} \yr$ old) is $\approx
0.16$, while that for a stellar disk (approximated by continuous star
formation over $10^9$ -- $10^{10} \yr$) is $\approx 0.13$--$0.14$.
Corrections for the approximation of red supergiant spectra by red
giant spectra are unlikely to increase the continuous star formation
CO index by more than $\sim 0.04$ at $10^9 \yr$ or by $\sim 0.02$ at
$10^{10} \yr$ (see figures 8--9 of Charlot \& Bruzual 1991, and tables
2--3 of Bell \& Briley 1991).  Thus, regions of a disk galaxy where
the star formation rate has not been unusually high in the preceding
$10^9 \yr$ need not have a CO index exceeding that for an elliptical
galaxy population.

\begin{figure}
\figurenum{3}
\epsscale{0.85625}
\plotone{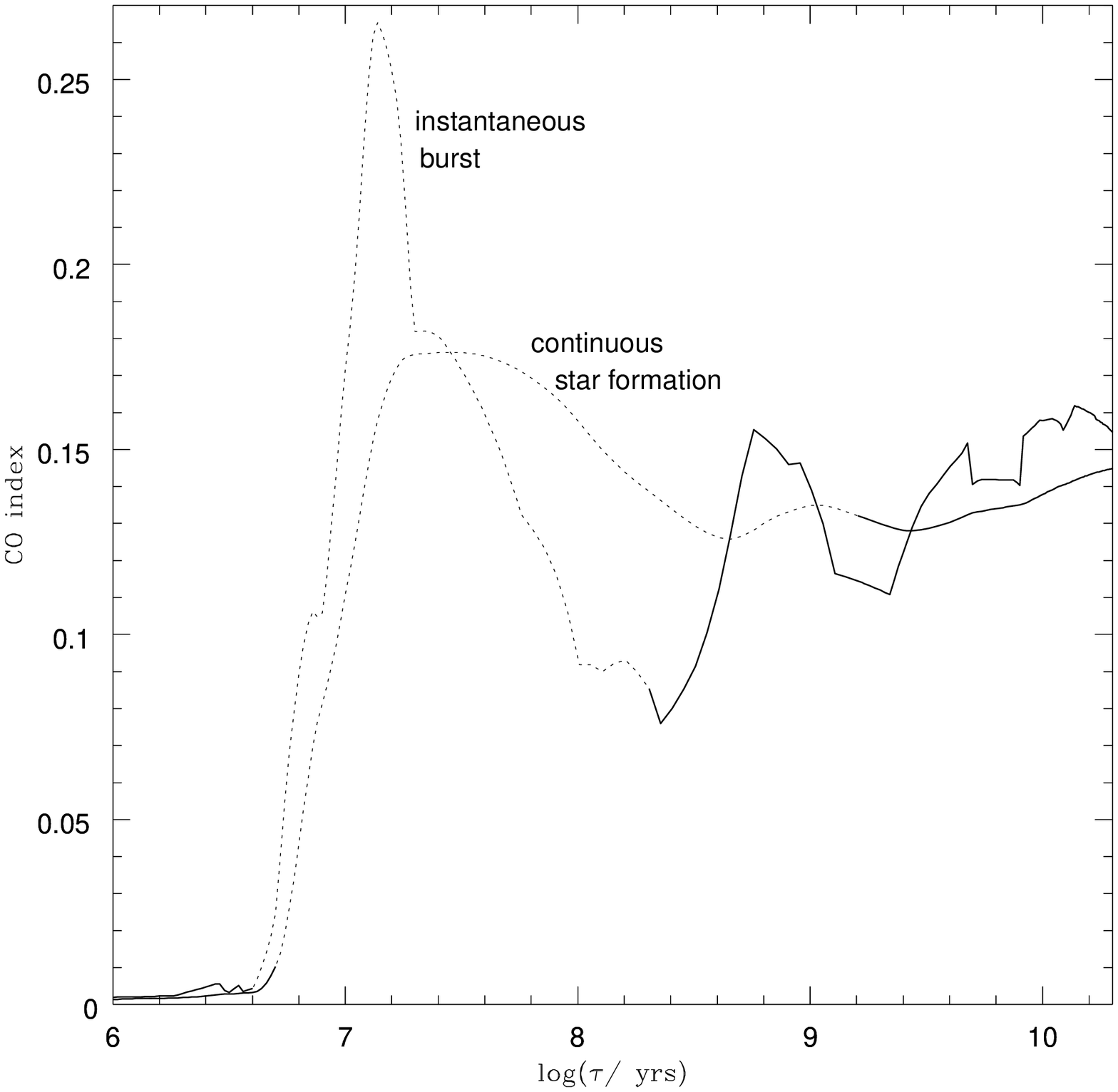}
\caption{
CO index as a function of stellar population age, based on a spectral
population synthesis code (``GISSEL'') by Bruzual \& Charlot (1993).
Curves for an instantaneous burst of star formation and for continuous
star formation at a fixed rate are shown.  We assumed a Scalo (1986)
initial mass function with cutoffs at $0.1 \Msun$ and $65 \Msun$.  For
regimes where red supergiants dominate the K band light, the model
indices will be underestimated, because spectrophotometric libraries
for red supergiants were not complete when the code was written, and M
giant spectra are used to represent the light of M supergiants.  The
CO index curves are dotted where supergiants contribute over half of
the total K band light, and solid elsewhere.  (The supergiant
contribution was estimated from figures 8 and 9 of Charlot \& Bruzual
1991.)
CO indices were computed by convolving model spectra with the Apache
Point Observatory $2.21 \micron$ and $2.34 \micron$ filters; the zero
point of the resulting CO indices was set to give zero CO index to the
model spectrum for Vega.  
}
\end{figure}

It is also possible that the comparison of our CO indices with those
of Frogel et al (1978) are affected by systematic differences between
the data sets.  One such difference could be in the filter system.
The zero point of our CO indices is constrained to match that of
Frogel et al by our choice of standard stars from the list of Elias et
al (1982).  However, the standard stars we observed all had small CO
indices, so that a multiplicative rescaling of our indices may be
required to bring them into agreement with the system of Frogel et al
(1978).  Differences in our K correction methods may also have a small
effect intercomparison of the indices.  

In any case, modest uncertainties in the absolute calibration of our
CO indices have comparatively minor effects on our analysis, which is
based primarily on {\it relative\/} CO indices between different
regions of disk galaxies.

\subsection{The fractional contribution of a young population}
We now wish to interpret spatial variations in the CO index in terms
of changes in stellar populations.  Consider a background stellar
population having CO index $B$, and a bright region with ongoing star
formation having index $S$.  Suppose that the difference in CO index
is due entirely to the presence of the young stellar population, and
that this population has CO index $Y$.  Together, $B$, $Y$, and $S$
determine the fractions $x_c$ and $x_b$ of the $2.21$ and $2.34
\micron$ light in the star forming region that must be due to the
young stellar population.  The relevant expressions (derived below) are
$$
x_c = {10^{0.4(S-B)} - 1 \over 10^{0.4(S-B)} - 10^{0.4(S-Y)} }
 \quad \hbox{\rm and} \quad
x_b = 10^{0.4(S-Y)} x_c = {10^{0.4(S-B)} - 1 \over 10^{0.4(Y-B)} - 1 } ~~.
$$
Assuming that the young population has metallicity at least that of
the older populations, and that its age is $\sim 10^7$ years so that
red supergiants are present, we expect $Y > B$.  $S$ will be
intermediate between $B$ and $Y$, because it is the index for a
mixture of the background population and the young population.
Under these conditions, the above relations may be approximated by
$$
x_c \approx x_b \approx (S-B) / (Y-B) ~~.
$$

To derive these results, we introduce auxiliary
quantities $\cB$, $\cS$, $\cY$, $\bB$, $\bS$, $\bY$, where 
 $c_i$, $b_i$ represent the flux densities from the
different stellar populations in the
bright patch at the continuum ($2.21 \micron$) and band head ($2.34
\micron$) filter bandpasses.  The subscripts $_B$, $_Y$, and $_S$
label the contributions from the background, young, and total
populations respectively, so that $ \cS = \cB + \cY$ and $ \bS = \bB +
\bY$.
By construction, $2.5 \log_{10}\left( \cB / \bB \right) + q = B$, so
that $(\cB / \bB) = 10^{0.4 (B - q)}$.  Similarly, $(\cY / \bY) =
10^{0.4 (Y - q)}$ and $(\cS / \bS) = 10^{0.4 (S - q)}$.  The
fractional contributions of the young stellar population to $\cS$ and
$\bS$ are then $x_c = \cY/\cS$ and $x_b = \bY / \bS$. Dividing these
gives 
$${x_b \over x_c} = {\bY \over \cY} {\cS \over \bS} =
10^{0.4(S-Y)} ~~,$$
and $ x_b = 10^{0.4(S-Y)} x_c $ as stated above.
The fractional contributions of the background population to $\cS$ and
$\bS$ are $1-x_c = \cB / \cS$ and $1-x_b = \bB / \bS$.  Dividing these
gives
 $${1 - x_b \over 1 - x_c} = {\bB \over \cB} {\cS \over \bS} =
10^{0.4(S-B)} ~~. $$
Substituting $10^{0.4(S-Y)} x_c$ for $x_b$ in
this expression and solving for $x_c$ gives the formulae stated above.

The largest tabulated CO indices in the models by Bell and Briley
(1991) are $\CO \approx 0.06 + 0.015[A/H]$ for $4500 K$ dwarfs, $\CO
\approx 0.21 + 0.06[A/H]$ for $4000 K$ giants, and $\CO
\approx 0.29 + 0.08[A/H]$ for $4000 K$ supergiants, where $[A/H]$ is
metallicity relative to solar
and is unlikely to
lie outside the range $0 \ga [A/H] \ga -1$ for the galaxies in our
sample.
The CO indices for composite stellar populations can attain values
$\sim 0.25$ (e.g., the strongest indices Persson et al [1983] measured
for star clusters in the Large Magellanic Cloud were $\CO = 0.255$ for
NGC~1767, $0.245$ for NGC~2002, and $0.24$ for NGC~1805).  This value
is also roughly consistent with the indices that \LRV96\ find in
population synthesis models for starbursts.  We will therefore use
$\CO = 0.25$ as an illustrative value of $Y$ in the following
discussion, while reminding the reader that $Y$ depends on the
starburst age, initial mass function, and metallicity in a complex
fashion.

Substituting in observed values for the bright patches in the NGC 278
ring ($15'' \la r \la 25''$), we find that
$S \approx 0.12$, while in the background regions at the same radius
$B \approx 0.06$.  Taking $Y = 0.25$, we then find $x_c \approx 0.33$,
i.e., one third of the
light must come from the extra stellar population.
This is consistent with the surface brightness increase in the bright
patches, which are a factor
$\sim 1.45$ above the background regions in the annulus $15'' \la r
\la 25''$.

In NGC 5713, the nuclear CO index ($0.17$) exceeds that in the nearest
background region ($\sim 0.10$).  If this is due to nuclear star
formation, our model with $S=0.17$, $B=0.10$, and $Y=0.25$ gives
$x_c \approx 0.48$, so the extra stellar population would have to contribute
about half the light.  The nuclear surface brightness exceeds that of
the disk at $r = 10''$ by a magnitude, so there is no difficulty here
in ``hiding'' the flux from an additional population.  However, a
young stellar population is not a unique explanation in this case,
because the metallicity may differ between the nucleus and the
comparison region, which is $\ga 1 \kpc$ away.

For NGC 2649, the typical background disk has a low CO
index, $B \sim 0.06$.  The upper limit to CO
index variations corresponds to $S - B \la 0.10$ for $r < 25''$.
Hence, we find $x \la 0.5$, i.e., the surface brightness is probably dominated
by older stars, although we cannot rule out a very substantial
contribution from young stars in this relatively low surface brightness system.

\subsection{Hypothesis testing}
We can consider our results in terms of two extreme null hypotheses:
First, that NIR light faithfully traces mass; and second, that the 
departures from some model are due entirely to the presence of star
formation activity.  The first implies that the CO index should not vary 
azimuthally, while such variations are expected in the second 
and can be modeled subject to certain assumptions.

Because the mass to light ratio of a stellar population is
sensitive to both age and metallicity, our first hypothesis in general
implies a spatially invariant CO index.  In practice, the dependence
of the K band mass to light ratio on metallicity is weak (e.g., Kodama
\& Arimoto 1997), so that a metallicity change of 1 dex can change the
CO index in an old population by $\sim 0.04$ (Bell \& Briley 1991) and
the K band mass to light ratio by $\sim 20\%$.  Thus, a radial
gradient in the CO index could be consistent with only a modest change
in mass to light ratio.  However, substantial {\em azimuthal}
gradients in metallicity are not expected in the disk component of a
galaxy, because phase mixing will suppress them within a few orbital
times.  Thus, azimuthal variations in CO index are best explained by
variations in the age structure of the stellar population.  The test
of our first null hypothesis is then to look for such azimuthal
variation.  NGC 278 clearly fails this test, while NGC 2649 and NGC
5713 pass given the uncertainties in the observations.

Producing a detailed prediction of the CO index as a function of
radius under the second hypothesis requires assumptions about the
stellar populations.  We prefer instead the more general model fitting
approach described in the following section.

\subsection{Model fitting}
In addition to null hypothesis testing, we can treat the difference
$\DCO$ between the CO indices of  background and bright regions as
a free parameter and determine its value through $\chi^2$ minimization.
The null hypotheses adopted above then become special cases of this more
general model with $\DCO = 0$ and $\DCO \approx 0.1$--$0.15$ respectively.
We have studied this model for several sets of CO index measurements,
corresponding to different regions of the sample galaxies.  The first of
these is the bar region in NGC 5713, defined as $8'' \le r \le 25''$,
where we find $\DCO = 0.035 \pm 0.010 [\pm 0.020]$.  (Here the first
error bar is derived from the photometric uncertainties in the data,
while the second is derived from bootstrap resampling of the input CO
index measurements.  The second error estimate thus represents the
uncertainty caused by spatial variations in $\DCO$.)  The second region
tested was the grand design spiral in NGC 2649 (all data with $r <
27''$), with $\DCO = 0.052 \pm 0.037 [\pm 0.053]$. 
The ``ring'' region of NGC 278 ($9'' < r < 30''$) shows $\DCO = 0.062 \pm
0.011 [\pm 0.038]$.
It thus appears that there is a marginal CO index excess in
the bar of NGC 5713, and a significant excess in the ring in
NGC 278 (although this absorption is itself patchy, leading to a large
error estimate from bootstrap resampling).   There could be absorption
in NGC 2649 but the data are not sensitive enough to make a
definitive statement.  The best fitting $\DCO$ yields $\chi^2$ per
degree of freedom of $1.4$ for NGC 2649, $2.7$ for NGC 5713, and $4.8$
for NGC 278.  This is quantitative evidence that the model is an
oversimplification, which is not surprising considering that variations in
star formation history and metallicity can result in a spatially
variable $\DCO$.

\subsection{The contribution of young populations to total luminosity}
We can estimate the contribution of red
supergiants to the total NIR light from the sample galaxies.  
For NGC 278, the bright regions in the two annuli with significant CO
index deviations account for total flux corresponding to
$K' = 10.48$, 
or $9 \%$ of the galaxy's total flux.  As shown above, $\sim$ one
third of this is from a young stellar population, so young stars
contribute not less than $\sim 3\%$ of the galaxy's total $2 \micron$
emission.
In NGC 2649, the bright region masks (here corresponding well to the
nucleus and spiral arm ridges) account for $53\%$ of the total light,
so no more than $1/4$ of the light can be from a young stellar
population.

Thus, while young stars do not dominate the global NIR flux
from disk galaxies, they can be locally quite important in star
forming regions.  This can bias arm:interarm amplitudes and similar
measures of galactic substructure.

\subsection{Aside: The distance to NGC 2649}
We can use our data together with published gas kinematics to estimate
the distance to NGC 2649 using the K band Tully-Fisher relation
(Malhotra, Spergel,
Rhoads, \& Li 1996).
The inclination angle of NGC 2649 can be estimated by assuming that a
logarithmic spiral describes its arm geometry. 
We found inclination angle $i = 30^\circ \pm 3^\circ$.  This can be
used to infer a deprojected line width.  The observed HI linewidth is
$\sim 260 \kms$ at 20\% of peak (Lewis 1987, quoted in Huchtmeier \&
Richter 1989).  The corrected width is then $\Delta V = 520 \pm 50
\kms$, somewhat exceeding the Milky Way's line width (cf. Malhotra et
al 1996) and implying an absolute K band magnitude $M_K = -24.0 \pm
0.4$.  The dominant uncertainty is from the inclination correction.  The
resulting distance is $d = 66 \pm 13 \Mpc$ (corresponding to $H_0 = 64
\pm 13 \kms / \Mpc$).  At this distance the
inferred scale length in our K band data is $5 \pm 1 \kpc$, larger
than the Milky Way (cf. Spergel, Malhotra,
\& Blitz 1996) and consistent with NGC 2649's larger rotation speed.

\section{Discussion}

We now wish to consider our results in the context of issues
raised in the introduction.

Interpreting measured CO indices in terms of the ages of stellar
populations is complicated by the variety of factors that can affect
CO absorption strength.  For individual stars, these factors are the
gravity, temperature, and metallicity at the photosphere.  These
factors in turn are primarily determined by the mass, initial
composition, and age of a star, though details of the star's history
(for example, binary interactions) may be important for some stars.
For a composite stellar population, the distribution of mass,
composition, and age can be derived from the initial mass function,
chemical evolution history, and star formation history of the galaxy.
In principle, then, we can determine the CO index expected from 
a given galactic evolutionary history, whence we can learn about the
contribution of recently formed stars to $2 \micron$ light.

In practice, the steps linking a model evolutionary history to an
inferred CO index encompass several uncertainties.  First, the problem
is underconstrained, so that additional information must be used.  In
practice this will entail assuming that the initial mass function (IMF) does
not vary with time and takes a form that fits observations of nearby
galaxies (e.g., Scalo 1986).  Second, the details of converting
stellar age, metallicity, and mass to a CO index are not well known.
Empirical approaches to this problem associate libraries of observed
stellar spectra with the controlling physical parameters.  A shortage
of well measured supergiant spectra in current libraries led Bruzual
and Charlot (1993) to use M giant spectra for M supergiants.
This will lead to an underestimate of
the supergiant contribution to the CO index.  The best available
tables of theoretical CO indices (Bell \& Briley 1991) do not extend to
effective temperatures $\Teff < 4000 \K$, and so do not
offer a complete solution to this difficulty.  Third, the late
evolution of massive stars is complicated by mass loss and convective
overshooting phenomena, and is thus difficult to predict on
theoretical grounds (Langer \& Maeder 1995).  This means that the
distribution of stellar age, mass, and metallicity can only
approximately be mapped into a distribution of luminosity and
effective temperature for the young, massive stars we are concerned
with.

Much careful work will be necessary to reduce the uncertainties in
population synthesis models of the CO index.  Many of the difficulties
can however be ignored if we study directly measured CO indices of
star clusters of known age.  In particular, the purely theoretical
difficulties of determining which stars become red supergiants and how
long they live in that phase go away.  Likewise, any incompleteness in
spectral libraries (whether theoretical or empirical) becomes
irrelevant.  Accurate age estimates for the observed clusters become
the main difficulty.  The uncertainties here are smaller
if the cluster ages are derived from main sequence fitting,
because we then depend on our understanding of main sequence stellar
evolution, which is more certain than our understanding
of supergiant phases of massive stars.

In figure~4, we plot CO indices (measured by Persson et al 1983)
against age estimates (by Elson \& Fall 1985, 1988) for star clusters
in the Large Magellanic Cloud (LMC).  Elson and Fall used main sequence
fitting to estimate some cluster ages and to calibrate an age sequence
in a two-color diagram; ages for the remaining clusters were determined
from their positions along this sequence.  The resulting CO-age
relation shows a statistically significant correlation with a large
scatter.  Besides the observational errors in CO index and age, this
scatter includes contributions from Poisson fluctuations in the
comparatively small number of evolved stars (especially in the younger
clusters; see Santos \& Frogel 1997) 
and from possible metallicity variations among clusters.
Metallicity differences between and within galaxies may also exist of course.
Comparing the CO indices for NGC 278 to those for LMC clusters yields
a characteristic age  $\sim 10^8$ years for the regions of
strongest CO absorption, and  $\sim 10^9$ years for the rest of the
disk.  In reality, of course, the CO index for diffuse light from a
galactic disk encompasses stars of a wide range of ages.  It is
nonetheless reassuring that the bright regions show a characteristic
age comparable to a typical galactic rotation time, while the
background disk appears much older than a rotation time.  Similarly, the
measured CO indices for NGC 2649 and for the background regions of NGC
5713 correspond to ages in excess of galactic rotation periods. 

\begin{figure}
\figurenum{4}
\plotone{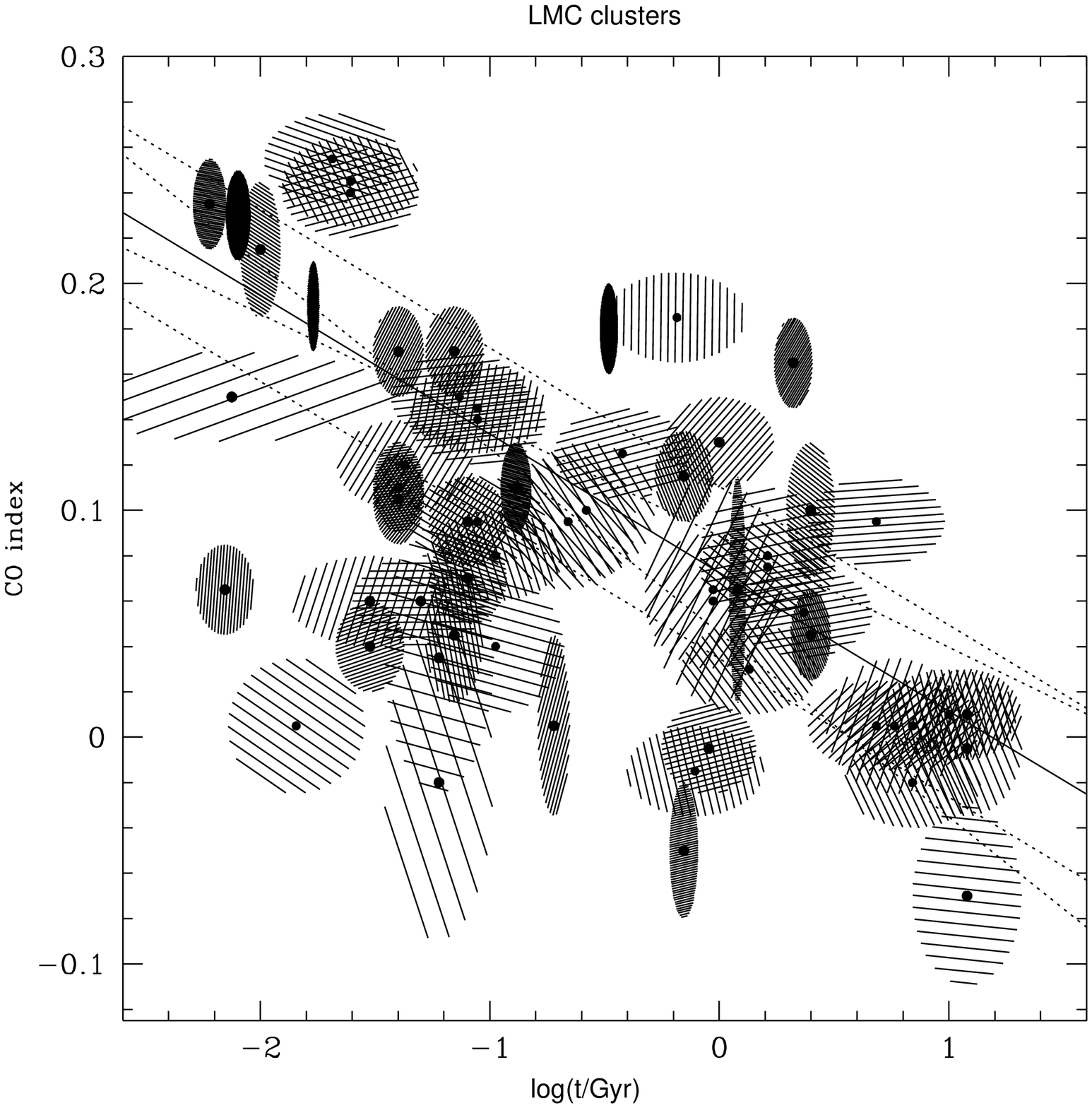}
\caption{
CO index as a function of age for star clusters in the Large
Magellanic Cloud.  60 clusters are shown.  Photometry is from Persson
et al (1983) and age estimates are from Elson \& Fall (1985, 1988).
The crosshatched regions are the $1 \sigma$ error ellipses for the
clusters.  Also shown are lines for the least squares fit (solid;
based on the 2D error ellipses shown) and for $1 \sigma$ changes in
the fitted line intercept and/or slope (dotted).}
\end{figure}

Our conclusions regarding the fraction of total NIR light that comes
from young stars are consistent with the small observed scatter in the
$2 \micron$ Tully-Fisher relation ($< 0.45 \mag$ at the 95\% level
[Malhotra et al 1996]).  If star formation rates differ widely among
galaxies of the same rotation speed, the scatter in the Tully-Fisher
relation cannot be smaller than the scatter in the fraction of light
from young stars.

Other indicators of star formation activity could also be used to
estimate the contribution of young stars to near-infrared light.  Each
involves its own set of uncertainties.  The CO bandhead is a direct
test for supergiant light in the $2 \micron$ band.  Other common
diagnostics trace either the presence of ultraviolet (UV) radiation
from hot, young stars, reprocessed either by gas (as recombination
lines) or by dust (as mid- and far-infrared emission); or the
byproducts of supernova shocks, reprocessed as synchrotron radiation.
The absence of such tracers could demonstrate that features in
near-infrared light are due to old stars.  However, their presence would not
unambiguously prove that NIR features are due to young stars, because
of the astrophysical uncertainties in determining the number of red
supergiants that evolve from a given population of young main sequence
stars.

\section{Conclusions}

We have measured the strength of the $2.3 \micron$ CO absorption bands
in three nearby disk galaxies.  The sample covers a
substantial range of morphologies and global star formation rates, but
does not include any starburst galaxies. 
We find that the CO index shows significant spatial variations in the
brightest regions of our two best-measured galaxies.  The variations
are consistent with $\sim 1/3$ of the light from these regions being
produced by a young stellar population whose near-infrared light is
dominated by red supergiant stars.  The characteristic age of such a
population, $\sim 10^7$ years, is smaller than galactic rotation
periods, so that these stars must still be near their birthplaces.
Globally, the contribution of such stars to the galaxies' near-IR
luminosity is a few percent.  Thus, while the overall $2 \micron$
emission is primarily due to old stellar populations, local features in the
near-infrared light of disk galaxies can be due primarily to young
stars.

\acknowledgments
I thank David Spergel, Jim Gunn, Michael Strauss,
Sangeeta Malhotra, Ed Fitzpatrick, Bohdan Paczy\'{n}ski, Stephane
Charlot, Neil Tyson, Hans-Walter Rix, and Jay Frogel (the referee)
for many helpful discussions during the course of this work.
Thanks also to Roger Bell, who kindly provided the model spectrum used
to calculate redshift corrections to the CO index.
I am also grateful to the Infrared Processing and Analysis
Center for its hospitality.  This research has been supported by NSF grant
AST 91-17388, NASA grant ADP NAG5-269, NSF traineeship
DGE-9354937, and a Kitt Peak National Observatory postdoctoral
fellowship.  This research has made use of the Simbad database,
operated at CDS, Strasbourg, France.  This research has made use of
the NASA/IPAC Extragalactic Database (NED) which is operated by the
Jet Propulsion Laboratory, California Institute of Technology, under
contract with the National Aeronautics and Space Administration.


\newpage

\figcaption{Images of the galaxies in $K'$ band ($2.12 \micron$)
and of the masks used for photometry.  Each image is 2.06 arcminutes on a side.
North is up, East is left.
Images:  a. NGC 2649; b. NGC 278; c. NGC 5713.
Masks: Black areas are bright region masks, grey areas
are background region masks. d. NGC 2649; e. NGC 278; f. NGC 5713.}

\end{document}